\documentclass[12pt]{article}
\usepackage{epsfig}
\usepackage{amssymb}

\def\be{\begin{equation}}
\def\ee{\end{equation}}
\def\ba{\begin{array}{c}}
\def\ea{\end{array}}

\newcommand{\bea}{\begin{eqnarray}}
\newcommand{\eea}{\end{eqnarray}}

\begin{document}

\begin{center}

{\Large \bf {


Avoided level crossings in quasi-exact approach

 }}

\vspace{20mm}

 {\bf Miloslav Znojil}

 \vspace{3mm}
Nuclear Physics Institute ASCR, Hlavn\'{\i} 130, 250 68 \v{R}e\v{z},
Czech Republic

\vspace{0.2cm}

 and

\vspace{0.2cm}

Department of Physics, Faculty of Science, University of Hradec
Kr\'{a}lov\'{e},

Rokitansk\'{e}ho 62, 50003 Hradec Kr\'{a}lov\'{e},
 Czech Republic

{e-mail: znojil@ujf.cas.cz}

\vspace{3mm}



\end{center}

\vspace{5mm}

\section*{Abstract}

In the quantum models described by analytic
potentials $V(r)$
with several pronounced minima
the phenomenon of tunneling
opens the
possibility of a sudden relocalization
of the system after
a minor modification of parameters.
This may really reorder the
minima and change, thoroughly,
the shape of the (measurable) probability density.
In the spectrum one observes
the ``avoided crossings''
of the energy levels.
By definition, the relocalization
configurations are
sensitive to perturbations
and
represent, therefore, a crossroad to alternative
evolution scenarios.
As long as
the numerical search for
these quantum analogues of the Thom's classical
catastrophes is not easy, a systematic non-numerical approach
is proposed here,
based on an
exact (or, better, quasi-exact) simultaneous construction
of mutually consistent
pairs of potentials $V(r)$ and wave functions $\psi(r)$.

\section*{keywords}

partially solvable
quantum models;
potentials with multiple minima;
relocalizations of probability densities;

\newpage

\section{Introduction}

One-dimensional Schr\"{o}dinger equation
 \be
 \left [-\frac{d^2}{d{r}^2}+V({r})
 \right ]\,\psi({r})=
 E\,\psi({r})\,,\ \ \ \ \psi({r})\in L^2(\mathbb{R})\,
 \label{se}
 \ee
often serves as a methodical laboratory and testing ground
guiding the analysis of various
more sophisticated (e.g., higher-dimensional or multi-particle)
models of the physical reality studied
on the level of microworld. Sometimes, it is being forgotten
that even the drastically simplified quantum model (\ref{se})
can also represent a system of its own, immediate physical interest.

In our present paper the latter idea is to be supported
by the turn of attention to the less elementary forms of
potentials $V(r)$
characterized by the presence of
several pronounced and deep,
more or less independent local minima
separated by non-negligible barriers
of a variable thickness.
In this sense our present study
was inspired by
our recent paper \cite{arnoldium}
in
which we revealed that in spite of an apparently purely
numerical character of any explicit
construction of predictions provided by a class
of general polynomial potentials, there exist several well formulated problems
(inspired by the classical Thom's  theory of catastrophes \cite{Zeeman})
in which the
predictions of certain
experimentally relevant ``quantum catastrophes''
could still be based on the use of certain
non-numerical approximation techniques.

In what follows a continuation of the latter project
will be
developed in two directions. First, in the context of
mathematics we shall
advocate the tractability of Eq.~(\ref{se})
even when potentials $V(r)$ possess certain
non-polynomial forms. Second, in the above-mentioned
context of physics of catastrophes modeled in a genuine quantum
meaning of the word,
we shall
propose and develop an innovative approach in which
one relocates, in some sense, the role of an input dynamical information
about the system from  $V(r)$ to a particular $\psi(r)$.

In section \ref{avolecro} we will outline
the basic physical motivation
of such a project. We will
explain the phenomenological as well as theoretical
usefulness of the specific experiment-related
concept of the so called avoided level crossing (ALC).
This outline
of motivation will continue in section~\ref{S2}
explaining
the merits of our proposal of analysis of
the ALC-related physics using
the mathematical methods
of construction of the so called
quasi-exactly solvable (QES) quantum models.

In section \ref{kedna} we will briefly
formulate our methodical message connecting the
exciting physical relevance of the deep multi-valley potentials
with an
optimality of the description of bound states
using the QES techniques.
The technical core of our paper
is then developed in section \ref{hudna}.
Via a detailed analysis of the QES models with
three or four almost separate
deep-valley subsystems we deduce and
support our main observation that
even a comparatively simple version of the QES
philosophy offers
a particularly fortunate interplay
between a desirable flexibility of the picture
of physics and a user-friendliness of its
mathematical representation using non-polynomial
but still elementary and analytic forms of potentials.

Section \ref{bedna} is then devoted to
the analysis of some specific consequences
of the tunneling through multiple barriers
and to the related approximate degeneracy of multiple
low-lying
states. We will point out, in particular, that
in such an arrangement
one has to deal with a very specific form of
the conventional oscillation theorems.
In subsequent section \ref{discussion} we
also turn attention
to a broader context of multiple-well models
and to their possible future role in the
quantum theory of catastrophes involving
more than just the relocalizations of ground states.

Section \ref{summary} is summary.


\section{Avoided level crossings\label{avolecro}}

The ubiquitous quantum phenomenon of avoided level crossings (ALC)
finds one of its most straightforward illustrations
in the one-dimensional
bound-state problem (\ref{se})
with a highly schematic rectangular double-well potential
 \be
 V(r)=V^{(RDW)}(r,a,b,c)=
 \left \{
 \begin{array}{ll}
 \infty\,,& r \in (-\infty,-3)\\
 a^2\,,& r \in (-3,-1)\\
 b^2\,,& r \in (-1,1)\\
 c^2\,,& r \in (1,3)\\
 \infty\,,& r \in (3,\infty)\,.
 \ea
 \right .
 \label{sqw}
 \ee
In the case of a high central barrier, $b^2 \gg \max (a^2,c^2)$,
the low-lying spectrum of energies $E_n$
with $n=0,1,\ldots, N_{\max}$
can be perceived as composed of the two
practically
independent
left-well and right-well low-lying approximate subspectra
 \be
 E_{p}^{\rm (left)}=a^2+\left [k_p^{\rm (left)}\right ]^2 \,,\ \ \ \
 k_p^{\rm (left)}=
 (p+1)\,\pi/2\,, \ \ p=0,1,\ldots, P_{\max} \,
 \label{Lenesqw}
 \ee
and
 \be
 E_{q}^{\rm (right)}=c^2+\left [k_q^{\rm (right)}\right ]^2 \,,\ \ \ \
 k_q^{\rm (right)}=
 (q+1)\,\pi/2\,, \ \ q=0,1,\ldots, Q_{\max}\,.
 \label{Renesqw}
 \ee
In such an extreme
dynamical regime
the wave functions will contain the well known \cite{Fluegge} single-well
components $\psi_{n_1}^{\rm (left)}(r)$ and $\psi_{n_2}^{\rm (right)}(r)$,
 \be
 \psi_{(n_1,n_2)}(r)=
 \left \{
 \begin{array}{ll}
 0\,,& r \in (-\infty,-3)\\
 \psi_{n_1}^{\rm (left)}(r)+ {\rm small \  corrections}\,,& r \in (-3,-1)\\
 0+ {\rm small \  corrections}\,,& r \in (-1,1)\\
 \psi_{n_2}^{\rm (right)}(r)+ {\rm small \  corrections}\,,& r \in (1,3)\\
 0\,,& r \in (3,\infty)\,.
 \ea
 \right .
 \label{psqw}
 \ee
In particular,
for the ground state with approximate energy
$E_0(a,c)=\min ( E_{0}^{\rm (left)}, E_{0}^{\rm (right)})=\min (a^2,c^2)+ \pi^2/4$,
we will have $\psi_0^{\rm (left/right)}(r) \sim -\cos k_0^{\rm (left/right)}r$
with $k_0^{\rm (left)}=\sqrt{E_0-a^2}$
and $k_0^{\rm (right)}=\sqrt{E_0-c^2}$.
In the generic cases
the subdominant
component will
be strongly suppressed
(i.e., $\psi_0^{\rm (right)}(r)$ will
lie too high at $a^2\ll c^2$,
and vice versa).
The related local suppressions of the probability density
will be an observable effect.

In the non-generic limit of $a^2\approx c^2$,
the ALC  effect
will occur. The approximate
coincidence of the
left-well and right-well subspectra
will be
accompanied by the approximate degeneracy
of the
first two
lowest energies, $E_0 \lessapprox E_1$.
The suppression
of the subdominant component of the wave function
will
temporarily disappear. Due to the process of an ALC-related
instantaneous exchange of dominance of the two subsystems, the
ground- and the first excited
state will form a doublet
characterized by different parities
but practically the same
probability density,
$ \psi^*_{1}(x)\psi_{1}(x) \approx \psi^*_{0}(x)\psi_{0}(x)$.

In our preceding paper \cite{arnoldium} we pointed out that
the price paid for the exact solvability of the
rectangular double- and multi-well
models as sampled by Eq.~(\ref{sqw}) is too high.
In any sufficiently realistic experimental setup
the shape of the potential is certainly different:
smooth and non-rectangular.
The more realistic polynomial
potential may still lead to a satisfactory approximate solvability
of the related Schr\"{o}dinger equation.
Moreover,
the polynomiality of potentials admits an analytic continuation
of the model.
This is a useful mathematical trick which can relate
the ALC-accompanying attraction/repulsion of energies to a proximity of the
so called Kato's exceptional-point (EP, \cite{Kato}) in
the complex plane of a suitable parameter \cite{EPappl}.

For all of these reasons
we proposed, in \cite{arnoldium}, that
the study of the ALC effects
might find a suitable benchmark-model background
in analytic Arnold-inspired polynomial
potentials $V_{(Arnold)}(r)$. This resulted in an
amendment of our understanding of some ALC-related
phenomena \cite{2D3D}. In parallel,
an important weakness of the innovation
may be seen in
a retreat from the exact solvability of
the rectangular-well models to the mere
approximate forms of the
solutions,
i.e., of the
verifiable and
measurable predictions.
This was a challenge which led to a
modification of the model-building strategy. Its form
will be described and illustrated in what follows.

\section{Conventional QES constructions\label{S2}}

\subsection{The change of paradigm}

The birth of the concept of
quasi-exactly solvable
(QES)
Schr\"{o}dinger equations
(see the compact
review
of its various aspects and, in particular, Appendix A
in Ushveridze's monograph \cite{Ushveridze})
contributed to an enhancement of efficiency of several
model-building strategies in quantum mechanics.
In the narrower context of Eq.~(\ref{se})
the essence of amendment
lies in a modification of
the conventional textbook philosophy
in which a ``known'' potential $V(r)$ is given
in advance while
the ``unknown''
bound states $\psi(r)$ must be reconstructed
via ordinary differential Schr\"{o}dinger Eq.~(\ref{se})
\cite{Fluegge}.
On this background
the QES-based model-building
strategy is more balanced,
transferring a part of the technical simplicity
assumptions from $V(r)$ to $\psi(r)$.

In our present application of the
QES model-building strategy
we decided to rewrite the above-mentioned Schr\"{o}dinger equation
into its formally equivalent form of definition of
a QES-compatible potential,
 \be
 V_{(QES)}({r})
 =E_{(QES)}+\psi''_{(QES)}({r})/\psi_{(QES)}({r})\,.
 \label{seb}
 \ee
In such an extreme version of the QES approach one
interchanges the roles of $V(r)$ and $\psi(r)$,
and one
inverts the conventional construction completely.
The duty of the carrier of the
physical input information is fully transferred from
$V(r)$ to an ansatz for a QES state $\psi(r)=\psi_{(QES)}({r})$.

Our
recommendation of replacement of Eq.~(\ref{se})
by Eq.~(\ref{seb})
found its immediate encouragement in
the emergence of certain descriptive
shortcomings of the poolynomial-interaction models as used
in \cite{arnoldium}.
A family of
Schr\"{o}dinger Eqs.~(\ref{se})
has been considered there
in a specific dynamical regime
controlled by
potentials
with pronounced multiple minima.
In this regime
the phenomenologically
most interesting
feature of the system
has been found to lie in the possibility of
a sudden change of the
topological structure of the probability densities.
We revealed that
in general, these changes
(called,  ``quantum
relocalization catastrophes'')
appeared caused by certain
very small changes of some parameters in the potential.

In \cite{arnoldium}, unfortunately,
the conventional
insistence on the elementary form
(viz., just polynomial form) of the
potentials implied that we could only work with
certain
approximate
forms of the wave functions.
The sensitivity of the observable catastrophic effects
to the physical parameters
interfered with the influence of the
round-off errors in $\psi(r)$.
This was a serious drawback and weakness of the method.
In our present paper we will show that
the unavoidable enhancement of the precision
of wave functions
can in fact be easily achieved
via the
less conventional QES approach.
In its framework, the original difficult
search for the instants of the
relocalization catastrophes based on the
brute-force solution
of Eq.~(\ref{se}) will drastically be simplified.

We should add that
the QES-based change of perspective is in fact much less
revolutionary than it may seem to be. Indeed,
from a strictly local point of view the difference
between ``the old exact input''  $V_{(Arnold)}(r)$ and
``the new exact input''  $\psi_{(QES)}(r)$ only becomes
visible on the level of corrections.
In particular,
potentials $V_{(Arnold)}(r)$ and $V_{(QES)}(r)$ are {\em both\,}
fairly well approximated, near their deep minima,
by the conventional and exactly solvable harmonic-oscillator wells.
For this reason, also the local
forms of the related ground-state wave functions
$\psi_{(QES)}(r)$ and  $\psi_{(Arnold)}(r)$
are both well approximated by the Gaussians.

\subsection{An illustrative sextic-anharmonic-oscillator example}

In the extensive dedicated
literature (cited, e.g., in \cite{Ushveridze})
interested readers could find several other,
less elementary
examples
and generalizations
of the QES idea and constructions.
Some of them possess an intriguing
mathematical background \cite{Turbiner}
while some other put more emphasis upon
nontrivial phenomenological
applications \cite{Shifman}.
Thus, for example, the most recent studies
are revealing new links between the QES models
and the theory of special functions \cite{Turbinerb}.
These samples of the technical and mathematical progress
are also currently accompanied by the innovative 
phenomenological applications of the various multi-well-shaped
potentials, say, in nuclear physics \cite{Turbinerc}.
For our present purposes, incidentally, we will only need the
technically less sophisticated version of the recipe.
In the overall QES context
the input information
represented exclusively by the potential
will be considered here
an expensive luxury.
A compensation
will be sought in a more balanced
model-building strategy.

One of the best known
and probably also one of the oldest
illustrations of such an attitude
is provided by the
sextic-polynomial interaction potential  
\cite{Turbinerb,Turbinerc,Singh}
 \be
 V^{}({r})=A\,{r}^2+B\,{r}^4+{r}^6\,.
 \label{shape}
 \ee
Its insertion in
Eq.~(\ref{se})
would only lead to a purely numerical problem in general.
In the QES setting,
therefore, the overall two-parametric definition (\ref{shape})
is to be simplified and accompanied
by an
additional requirement that
the
underlying Schr\"{o}dinger
equation should generate at least one
wave function which is exact and elementary.
Once such a function is
prescribed, say, in the simplest possible
one-parametric single-exponential ground-state form
 \be
 \psi_{(QES)}({r})= \exp \left [-\frac{1}{4}({r}^2+\alpha)^2 \right ]\,
 \label{solu}
 \ee
it is still possible to guarantee that
our Schr\"{o}dinger Eq.~(\ref{se}) will be
satisfied. Indeed,
it is an entirely
elementary exercise to verify that
for the specific QES ansatz (\ref{solu})
it is sufficient to
reparametrize, self-consistently, the
energy   $E=\alpha$ in Eq.~(\ref{se}) as well as the
couplings
$A=A(\alpha)=\alpha^2-3\,$ and $B=B(\alpha)=2\alpha$
in Eq.~(\ref{shape}).


Besides the methodical appeal of
ansatz (\ref{solu}) also
the related picture of
dynamics
is remarkable. The
QES version of potential (\ref{shape})
acquires the single-well shape at $\alpha>\sqrt{3}$,
the double-well shape at $-\sqrt{3}<\alpha<\sqrt{3}$, and
the triple-well shape at $\alpha<-\sqrt{3}$.
Unfortunately,
the variability of the shape of $V({r})$
is only partially paralleled by the flexibility of
the measurable probability density $\varrho(x) = \psi^*(x)\psi(x)$
possessing just two maxima at most.

The latter observation
gave birth to our present study.
In essence,
a phenomenologically richer menu of bound states
will be obtained via a multi-term extension of the class of the
QES ansatzs as sampled by Eq.~(\ref{solu}).


\section{Relocalization catastrophes\label{kedna}}

In the theory of classical dynamical systems
the concept of equilibrium
can be given
a neat geometric interpretation
\cite{Zeeman}. The related mathematics also offers
a systematic
classification of the
processes during which these equilibria
are being lost or established \cite{Arnold}.
These considerations
were
made particularly popular
by Ren\'{e}
Thom
\cite{Thom}
who selected seven most elementary
transmutations of the classical equilibria
and gave them specific nicknames like ``cusp'', etc.

Whenever one tries to extend the Thom's systematics
to the theory of quantum dynamical systems
one must imagine
that in quantum dynamical systems
the predictions of any measurable effect become
merely
probabilistic, based on Schr\"{o}dinger equation.

\subsection{Polynomial multi-barrier potentials\label{kapje}}

In a way inspired by the Arnold's treatment of the
{\em classical\,} catastrophe theory
\cite{Arnold}) we considered, in our recent paper \cite{arnoldium},
all of the confining and
spatially-symmetric polynomial {\em quantum} potentials
 \be
 V_{(Arnold)}(r) = x^{2N+2}+c_1\,r^{2N} + c_2\,r^{2N-2}
 + \ldots + c_{N}\,r^2\,,\ \ \ \ r \in (-\infty,\infty).
 \label{arno}
 \ee
We restricted our attention to the multi-well
dynamical regime in which the
potential developed an $N-$plet of
high and thick barriers.
Near a pronounced and dominant
minimum of the potential (i.e.,
say, at $r \approx a$ such that
$V'_{(Arnold)}(a)=0$)
the ground state energy
proves well approximated
by
the leading-order formula
  \be
 E_0 \approx V_{(Arnold)}(a)+\omega^{(\min)}
 \,,\ \ \ \
 \omega^{(\min)} =
 V''_{(Arnold)}(a)/2\,.
 \label{leading}
 \ee
Naturally, under an additional, {\it ad hoc\,}
assumption of
the left-right symmetry of the Arnold's potential (\ref{arno})
we have to keep in mind that
unless $a=0$ these global minima
occur in pairs.
This means that
the approximate form of the generic ground-state
wave function will have two components at $a\neq 0$,
 \be
 \psi^{(Arnold)}(r) \sim
  \exp[-\omega^{(\min)}(r-a)^2]
  +\exp[-\omega^{(\min)}(r+a)^2]\,
  \label{sido}
 \ee
Even if the
anharmonic corrections to
the generic formula (\ref{sido}) remain negligible
one should take into consideration
also all of the non-generic situations in which
the
nontrivial
contributions to $\psi^{(Arnold)}(r)$
come
from the broader subdominant
local minima $r =\pm a_j$
of the potential
satisfying the condition of coincidence of the
leading-order ground-state
energies,
 \be
 E=E_0(j) = V_{(Arnold)}(a_j)+\omega^{(\min)}_j=E(M)\,,
 \ \ \ \ j=1,2,\ldots,M\,.
 \label{tydomi}
 \ee
In this setting
the reliability of
the leading-order approximation
must be based not only on a guarantee of the
sufficiently large
size of all of the constants $\omega^{(\min)}_j$
with $j=1,2,\ldots,M$
(making the
first local excitations
$E_1(j)=E_0(j)+2\,\omega^{(\min)}_j$
sufficiently well separated)
but, first of all,
on a sufficiently strong suppression of the
potential influence of the ground-state
contributions
coming from all of the other potentially eligible local minima,
 \be
 E_0(k) = V_{(Arnold)}(a_k)+\omega^{(\min)}_k
 \gg E(M)\,,
 \ \ \ \ k=M+1, M+2, \ldots, N+1\,.
 \label{nedom}
 \ee
In
such a case the leading-order
ground-state wave functions will
have the form of superpositions
 \be
 \psi^{(Arnold)}_0(r) \sim
 \sum_{j=1}^M\, p_j\,\left [
 \exp[-\omega^{(\min)}_j(r-a_j)^2]
 +
 \exp[-\omega^{(\min)}_j(r+a_j)^2]
 \right ]\,.
  \label{dosic}
 \ee
In \cite{arnoldium} it has been emphasized that the latter solutions
will be highly sensitive to certain very small changes of parameters $c_k$
in potentials (\ref{arno}).
Such a change will cause, typically, a
relocation of a
subset of the $M-$plet
of the initial local minima from the dominant category (\ref{tydomi})
to the subdominant category (\ref{nedom}). In Eq.~(\ref{dosic}),
due to the tunneling,
the respective coefficients $p_j$ will then vanish. For this reason
also the related probability density
$\varrho(x) = \psi^*(x)\psi(x)$
will change accordingly.
The process (during which the value of $M$ decreased)
can be also inverted (leading to an increase of
the number $M$ of participating local minima).
Finally, a combination of the two processes
(initiated and ending, in long run, at the two different
and stable equilibria with $M=1$) has been given, in \cite{arnoldium},
the name of a relocalization quantum catastrophe.
In this context, our present main ambition may be
briefly characterized as
a QES-based closed-form description of the relocalization-catastrophe
instants with an arbitrarily large degree of
the instantaneous degeneracy $M>1$.


\subsection{Double-well Gaussian ansatz}

In a certain parallel to the above-outlined non-smooth model
(\ref{sqw}) let us now contemplate the
smooth and most elementary pair-of-oscillators ansatz
 $$
 \psi^{(2)}(r)=\exp[-(r-a)^2]+\exp[-(r+a)^2]\,.
 $$
At a large separation parameter $a\gg 1$ it represents,
locally, two independent harmonic oscillators
in ground state.
Such an asymptotically elementary formula offers a perfect insight
in the shape of the wave function
but as a starting point of
reconstruction of a related potential
it looks clumsy and can be simplified,
 \be
 \psi^{(2)}(r)=2\,\exp(-r^2-a^2)\,{\rm cosh}\,(2ar)\,.
  \ee
This function is better suited for insertion
in formula (\ref{seb}) yielding
 $$
 V({r})=E+\psi''(r)/\psi(r)=E-2
 +{4a^2}
 +4\,r^2
 -8\,a r\,{{\rm tanh}\,(2ar)}\,.
 $$
As long as the anharmonicity remains asymptotically subdominant
we may immediately conclude that the high excitations
will feel it as a not too essential perturbation.
The situation becomes different
for the low-lying bound states because
the potential acquires, near the origin, the double-well shape at
$a^2>1/4$.
Once we set, say, $V(0)=0$,
the explicit ground-state QES energy value
becomes equal to $E_0=2-{4a^2}$.
Its value decreases with the growth of $a^2$
and it drops below the local maximum of the potential at
$a^2=1/2$.
Subsequently, the potential
acquires the clear double-well shape.
At the sufficiently
large $a \gg 1$ we get $V(r) \approx 4\,r^2 -8\,a |r|$, i.e.,
$V(r) \approx 4\,(r-a)^2-4\,a^2$ at the positive $r \gg 1$, and
$V(r) \approx 4\,(r+a)^2-4\,a^2$ at the negative $r \ll -1$.
The minima at $r \approx \pm a$ are deep so that
with good precision
the low-lying spectrum becomes tractable as a
well-separated pair of the two remote harmonic oscillators.

A generalization of this observation will be
slightly more complicated but straightforward.

\section{Multi-Gaussian models\label{hudna}}

The QES approach
based on Eq.~(\ref{seb})
makes it clear that the conventional constraints of the polynomiality
of $V(r)$ limit the variability of the
shapes of the wave functions
$\psi(r)$. In \cite{arnoldium}, the necessarily
non-exact, approximate forms of the wave function solutions
led also to the mere approximate form of the predictions
of the measurable effects.
Another technical
obstruction consisting in
the non-geometric, deeply quantum nature of the
relocalization
catastrophes
has been also circumvented
using suitable approximations
of the energies as sampled by formula (\ref{leading}) above.

In this sense we
were only able to rely on approximate results. In the present paper, in contrast,
we will insist on the exact QES form of the wave functions.
This will enable us to
move to the more complicated QES ansatzs and, ultimately, to
oppose, constructively, the not too deeply motivated traditional
restriction of the form of the dynamical input information
to the mere polynomial class of the potentials.

\subsection{Three equidistant Gaussians \label{431}}

The $M=3$ QES ansatz
 $$
 \psi^{(3)}(r)=\exp[-(r-b)^2]+\exp(-r^2)+\exp[-(r+b)^2]
 $$
can be simplified,
 \be
 \psi^{(3)}(r)=\exp(-r^2)
 [1+2\,e^{-b^2}{\rm cosh}\,(2br)
 ]
 \label{431eq}
 \ee
and factorized,
 $$
 \psi(r)=
{e^{-{r}^{2}}}\,F( {}\,r )\,.
 $$
On this ground we easily evaluate
 $$
 \psi'(r)=-2\,r\,\psi+
{e^{-{r}^{2}}}\,F'( {}\,r )
 $$
and
 $$
 \psi''(r)=(4\,r^2-2)\,\psi-4\,r\,{e^{-{r}^{2}}}\,F'( {}\,r )+
{e^{-{r}^{2}}}\,F''( {}\,r )
 $$
so that, finally,
 \be
 V({r})=E+\psi''(r)/\psi(r)=E-2+4\,r^2
 -4\,r\,\frac{F'( {}\,r )}{F( {}\,r )}
 +\frac{F''( {}\,r )}{F( {}\,r )}\,.
 \label{ele}
 \ee
Although such a QES potential is not a polynomial,
its form remains elementary in general.
In particular, at $M=3$ we have
 $$
 F(r)=1+2\,e^{-b^2}{\rm cosh}\,(2br)
 $$
so that
 $$
 F'(r)=4b\,e^{-b^2}{\rm sinh}\,(2br)
 $$
with
 $$
 F'(r)/F(r)=\frac{4b\,e^{-b^2}{\rm sinh}\,(2br)}{1+2\,e^{-b^2}{\rm cosh}\,(2br)}
 $$
and with
 $$
 F''(r)=8b^2\,e^{-b^2}{\rm cosh}\,(2br)
 $$
in
 $$
 F''(r)/F(r)=\frac{b^2\,(8e^{-b^2}{\rm cosh}\,(2br)+4-4)}{1+2\,e^{-b^2}{\rm cosh}\,(2br)}=4b^2-\frac{4b^2\,}{1+2\,e^{-b^2}{\rm cosh}\,(2br)}\,.
 $$
This yields the potential
 \be
  V(r)-E_0 =-2+4\,b^2+4r^2+A/B\,,\ \ \ \
  \label{poa4}
  \ee
where
  $$A=
 { {-16\,br\sinh
\left( 2\,br \right) {e^{-{b}^{2}}}-4\,b^2}}\,,\ \ \ \
B ={2\,\cosh \left( 2\,br \right)
{e^{-{b}^{2}}}+1}\,.
 $$

%
%
%
%
%
%
%
%
\begin{figure}[h]                     
\begin{center}                         
\epsfig{file=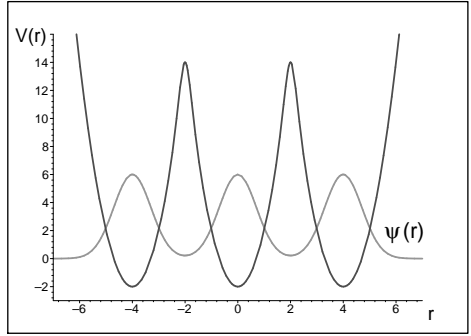,angle=270,width=0.35\textwidth}
\end{center}                         
\vspace{-2mm}\caption{The choice of a sufficiently large
shift in the
QES ansatz (\ref{431eq}) (here, $b=4$) keeps
the barriers sufficiently thick to suppress the tunneling
between the three deep wells (in our units, the display of
the QES input wave function is six times enlarged).
 \label{fia4}}
\end{figure}

Although the latter algebraic presentation of the resulting QES potential
is exact, it is both unusual and rather counterintuitive.
Its better perception can easily be obtained when we choose
any sufficiently large value of the parameter (say, $b=4$)
and draw a picture (see Fig. \ref{fia4}).
Its inspection immediately reveals that the shape of
potential (\ref{poa4}) is in fact just a chain of four
independent
wells of an approximately harmonic-oscillator form.

\begin{figure}[h]                     
\begin{center}                         
\epsfig{file=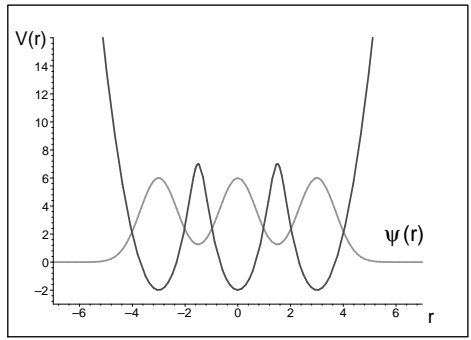,angle=270,width=0.35\textwidth}
\end{center}                         
\vspace{-2mm}\caption{A perceivable thinning of the barriers
followed by an enhancement of the tunneling
occurs at an intermediate shift $b=3$ in the
QES ansatz (\ref{431eq}).
 \label{fia3}}
\end{figure}

Once we choose a smaller but still sufficiently large value of $b=3$,
the overall picture does not change too much (see Fig.~\ref{fia3}).
Only after we further move to the still smaller value of $b=2$
(see Fig.~\ref{fia2}), we discover what happens when the overlaps
between the Gaussians become large.
At a still smaller $b=\sqrt{2}$ the wave function itself
even ceases to possess
the local minima although the
triple-well structure of the potential still survives.
At $b=1$ one only finds a not too pronounced
double-well shape of $V(r)$, and there are no traces of the shift
left in the potential at $b=1/2$.

\begin{figure}[h]                     
\begin{center}                         
\epsfig{file=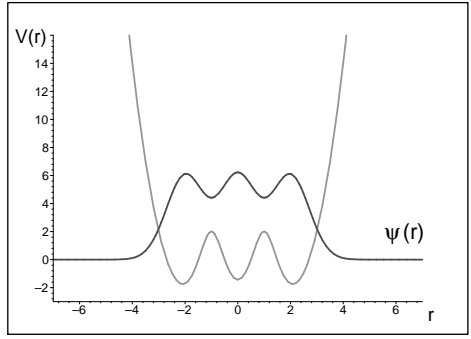,angle=270,width=0.35\textwidth}
\end{center}                         
\vspace{-2mm}\caption{At the small shift $b=2$ in (\ref{431eq})
the barriers are too low to suppress the tunneling.
 \label{fia2}}
\end{figure}

All of the
expectations based on the
pictures may be complemented by the easy but
rigorous Taylor-series expansions of the potential at any
relevant coordinate of interest. For example, in the vicinity of
the origin we get, at an arbitrary parameter $b$, formula
 $$
 V(r)-E=2\,{\frac
 {4\,{b}^{2}{e^{-{b}^{2}}}-2\,{e^{-{b}^{2}}}-1}{2\,{e^{-{b}^{2}}}+1}}+
 $$
 $$
 +4\,{\frac {-16\,{b}^{2}{e^{-2\,{b}^{2}}}-8\,{b}^{2}{e^{-{b}^{2}}}
 +4\,{e^{-2\,{b}^{2}}}+4\,{e^{-{b}^{2}}}+1
 +4\,{b}^{4}{e^{-{b}^{2}}}}{ \left( 2\,{e^{-{b}^{2}}}+1 \right)
 ^{2}}}{r}^{2}-
 $$
 $$
 -16/3\,{\frac{{b} ^{4}{e^{-{b}^{2}}}
 \left(-32\,{e^{-2\,{b}^{2}}}-8\,{e^{-{b}^{2}}}+4
 +10\,{b}^{2}{e^{-{b}^{2}}}-{b}^{2} \right) }
 { \left(2\,{e^{-{b}^{2}}}+ 1 \right) ^{3}}}{r}^{4}+O \left( {r}^{6} \right)\,,
 $$
etc. For comparison, we can expand the wave function,
 $$
 \psi(r)=
 (2\,{e^{-{b}^{2}}}+1+ \left( 4\,{b}^{2}{e^{-{b}^{2}}}-2\,{e^{-{b}^{2}}
}-1 \right) {r}^{2}+ \left(
4/3\,{b}^{4}{e^{-{b}^{2}}}+{e^{-{b}^{2}}}+ {\frac
{1}{2}}-4\,{b}^{2}{e^{-{b}^{2}}} \right) {r}^{4}+O \left( {r}^{ 6}
\right) )
$$
and also its derivative,
 $$
 \psi'(r)=\left( 8\,{b}^{2}{e^{-{b}^{2}}}-4\,{e^{-{b}^{2}}}-2 \right) r
 + \left( -16\,{b}^{2}{e^{-{b}^{2}}}+4\,{e^{-{b}^{2}}}+2
 +16/3\,{b}^{4}{e^{-{b}^{2}}} \right) {r}^{3}
 +
 $$
 $$
 +\left(-8\,{b}^{4}{e^{-{b}^{2}}}
 -2\,{e^{-{b}^{2}}}-1+12\,{b}^{2}{e^{-{b}^{2}}}
 +{\frac{16}{15}}\,{b}^{6}{e^{-{ a}^{2}}} \right) {r}^{5}+O \left( {r}^{6}
 \right)\,
 $$
serving, e.g.,  the purposes of a more detailed analysis.

%
%
%
%
\begin{figure}[h]                     
\begin{center}                         
\epsfig{file=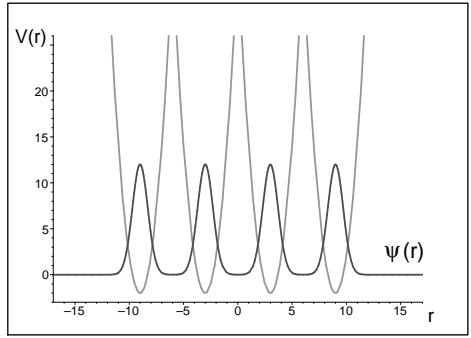,angle=270,width=0.35\textwidth}
\end{center}                         
\vspace{-2mm}\caption{At the value of  $a=3$ in (\ref{63eq})
the three barriers prove thick enough to suppress the tunneling.
 \label{fi4a3}}
\end{figure}

\subsection{Four equidistant Gaussians}

Once we wish to reconfirm the above picture- and intuition-based messages
we may just repeat the construction at $M=4$, with the QES ansatz
 $$
 \psi^{(4)}(r)=\exp[-(r-3\,a)^2]+\exp[-(r-a)^2]+\exp[-(r+a)^2]+\exp[-(r+3\,a)^2]
 $$
abbreviated as
 $$
 \psi^{(4)}(r)=2\,\exp(-r^2-a^2)\,[{\rm cosh}\,(2ar)+e^{-8a^2}{\rm cosh}\,(6ar)
 ]
  $$
and yielding
 $$
 F(r)={\rm cosh}\,(2ar)+e^{-8a^2}{\rm cosh}\,(6ar)\,,
  $$
 $$
 F'(r)=2a\,{\rm sinh}\,(2ar)+6a\,e^{-8a^2}{\rm sinh}\,(6ar)\,
  $$
and
 $$
 F''(r)=4a^2\,{\rm cosh}\,(2ar)+36\,a^2\,e^{-8a^2}{\rm cosh}\,(6ar)\,.
  $$

\begin{figure}[h]                     
\begin{center}                         
\epsfig{file=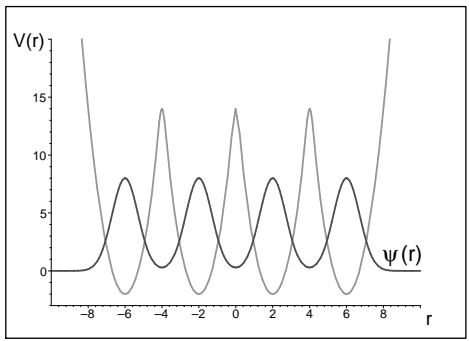,angle=270,width=0.35\textwidth}
\end{center}                         
\vspace{-2mm}\caption{For ansatz (\ref{63eq})
the lowering of the barriers and
the first marks of the tunneling emerge
at $a=2$.
 \label{fi4a2}}
\end{figure}

The existence of the quadruplet of the coordinates $ -3 a, -a, a, 3a $
approximating the local minima of $V(r)$
could again be shown to lead to the analogous graphical results
at $a=3$ (see Fig.~\ref{fi4a3}) and at $a=2$ (see Fig.~\ref{fi4a2}).
In the latter picture with $a=2$
we see again an onset of the tunneling
which becomes detected but remains weak.
In contrast,
at $a=1$ one would already get a strong
tunneling supporting a four-bump analogue of
the
three-bump structures as depicted in Fig.~\ref{fia2} above.
With the further decrease of $a$
the wave functions and, slightly later,
also the potentials would lose again
their local minima and maxima.
Only the potential themselves still exhibit,
even at a very small shift $a=1/2$,
the very week remnants
of their last two minima.

In the language of algebra with
\be
\psi(r)={e^{-{r}^{2}}} \left( 2\,\cosh \left( 2\,ar \right)
{e^{-{a}^{2}}}+2\, \cosh \left( 6\,ar \right) {e^{-9\,{a}^{2}}}
\right)
\label{63eq}
\ee
we get the numerator of the potential
in the form
 $$
 4\,{e^{-{a}^{2}-{r}^{2}}} ( -\cosh \left( 2\,ar \right)
 -\cosh
 \left( 6\,ar \right) {e^{-8\,{a}^{2}}}+2\,{r}^{2}\cosh \left( 2\,ar
 \right) +2\,{r}^{2}\cosh \left( 6\,ar \right) {e^{-8\,{a}^{2}}}-
 $$
 $$
 -4\,r
 \sinh \left( 2\,ar \right) a-12\,r\sinh \left( 6\,ar \right)
 a{e^{-8\, {a}^{2}}}+2\,\cosh \left( 2\,ar \right) {a}^{2}+18\,\cosh
 \left( 6\,ar
 \right) {a}^{2}{e^{-8\,{a}^{2}}} )
 $$
so that the ultimate formula for $ V(r)-E$
%
could again be analyzed in a routine manner.


\section{Asymptotic degeneracy
\label{bedna}}

\subsection{Higher numbers of barriers}

The comparison of the series of Figs. \ref{fia4} - \ref{fia2}
and \ref{fi4a3} - \ref{fi4a2} demonstrates that the
QES models generated by the three and four
equidistant Gaussians in ansatz $\psi_{(QES)}(r)$
share all of the relevant qualitative features of the
shape of the potential. In particular,
whenever the distance between Gaussians
(i.e., parameter $a$ or $b$)
exceeds a critical
(and, in fact, not too large) value,
one can perceive the QES potential as
a multiplet of several well separated harmonic oscillators.
As long as the tunneling through the
thick and high repulsive barriers becomes negligible,
the QES construction can immediately be extended
to the
general scenarios
in which the potential becomes composed
of $M$ wells separated by $M-1$ barriers
at any $M$.

In the algebraic language this means that
besides the above-listed $M \leq 4$ cases
we can also work with the next few ansatzs
 $$
 \psi^{(5)}(r)=\exp[-(r-2\,b)^2]+\exp[-(r-b)^2]+\exp(-r^2)
 +\exp[-(r+b)^2]+\exp[-(r+2\,b)^2]\,,
 $$
 $$
 \psi^{(6)}(r)=\exp[-(r-5\,a)^2]+\exp[-(r-3\,a)^2]+\exp[-(r-a)^2]+
 $$
 $$
 +\exp[-(r+a)^2]+\exp[-(r+3\,a)^2]+\exp[-(r+5\,a)^2]\,,
 $$
 $$
 \psi^{(7)}(r)=\exp[-(r-3\,b)^2]+\exp[-(r-2\,b)^2]+\exp[-(r-b)^2]+\exp(-r^2)
 +
 $$
 $$
 +\exp[-(r+b)^2]+\exp[-(r+2\,b)^2]+\exp[-(r+3\,b)^2]\,
 $$
etc. For the purposes of the QES construction
it makes sense to simplify
 $$
 \psi^{(5)}(r)=\exp(-r^2)
 [1+2\,e^{-b^2}{\rm cosh}\,(2br)+2\,e^{-4\,b^2}{\rm cosh}\,(4br) ]\,,
 $$
 $$
 \psi^{(6)}(r)=2\,\exp(-r^2-a^2)\,[{\rm cosh}\,(2ar)+e^{-8a^2}{\rm cosh}\,(6ar)
 +e^{-24a^2}{\rm cosh}\,(10ar)]\,,
  $$
 $$
 \psi^{(7)}(r)=\exp(-r^2)
 [1+2\,e^{-b^2}{\rm cosh}\,(2br)+2\,e^{-4\,b^2}{\rm cosh}\,(4br)
 +2\,e^{-9\,b^2}{\rm cosh}\,(6br)]
 $$
etc. With the growth of $M$ these superpositions of Gaussians
lead to the more complicated potentials but it is easy to see that
all of these potentials remain finite and comparatively elementary.

%

\begin{table}[h]
\caption{Shapes and zeros of
$\psi_n^{(3)}(r)$
at large $a \gg 1$
[symbols explained in the text].
}
\vspace{0.5cm}
 \label{owe3}
\centering
\begin{tabular}{||r||c|c|c|c|c||}
\hline\hline
 & $r \approx$ $-a$& $r \approx$ $-a/2$&
 $r \approx$ $ 0$& $r \approx$ $+a/2$& $r \approx$ $+a$\\
\hline
 \hline
 $n=0$& $\bigcap$ &$\smallsmile$&$\bigcap$&$\smallsmile$&$\bigcap$\\
 \hline
 $n=1$& $\bigcap$ &$\backsim$&$\bullet$&$\backsim$&$\bigcup$\\
 \hline
 \vspace{-.4cm}
 &$\bigcap$\ &$\backsim$&&$\sim$&$\bigcap$\\
 $n=2$& &$\bullet$&$\bigcup$&$\bullet$&\\ \hline \hline
\end{tabular}
\end{table}

\subsection{Low-lying excited states}

The locally bounded rational-function form of the QES potentials
(\ref{ele}) admits the routine
numerical construction of any excited non-QES bound state
by means of the brute-force solution of Eq.~(\ref{se}).
With the growth of the number
of barriers
such a solution becomes less and less comfortable.
At the same time,
the overall picture of physics remains simplified
because the
growth of the barriers
makes the separate
local minima of the
potential comparatively independent.
For this reason,
the double-well structure
of the global potential with $M=2$ will support
an almost degenerate doublet of bound states formed by the spatially
symmetric (i.e., nodeless) ground state and by the spatially antisymmetric
first excited state (with the single nodal zero at $r=0$). At the same time,
the second and third excited states will form
another clearly separated but again almost degenerate doublet, etc.

\begin{table}[h]
\caption{Shapes and zeros of the $M-$plet of
$\psi_n^{(M)}(r)$
at $M=4$.
 }
\vspace{0.5cm}
 \label{owe4}
\centering {\small
\begin{tabular}{||r||c|c|c|c|c|c|c||}
      \hline \hline
 $\ \ \ \ \ \ r \approx $& $-3b$&$-2b$&$-b$&$\ 0\ $
  &$+b$&$+2b$&$+3b$\\
  \hline
 \hline
 $n=0\ \ \ \ \ \ $& $\bigcap$&$\smallsmile$&$\bigcap$&$\smallsmile$
  &$\bigcap$&$\smallsmile$&$\bigcap$\\
  \hline
 \hline
 \vspace{-.4cm}
 & $\bigcap$&$\smallsmile$&$\bigcap$&$\backsim$
 && &\\
 $n=1\ \ \ \ \ \ $&&&&$\bullet$&$\bigcup$&$\smallfrown$&$\bigcup$\\
     \hline
     \hline
 \vspace{-.4cm}
 &$\bigcap$&$\backsim$&&&&$\sim$ &$\bigcap$\\
  $n=2\ \ \ \ \ \ $& &$\bullet$&$\bigcup$&$\smallfrown$&$\bigcup$&$\bullet$
     &\\
\hline
  \hline
 \vspace{-.4cm}
 &$\bigcap$&$\backsim$&&$\sim$&$\bigcap$&$\backsim$ &\\
 $n=3\ \ \ \ \ \ $& &$\bullet$&$\bigcup$&$\bullet$&&$\bullet$
     &$\bigcup$\\
       \hline \hline
\end{tabular}}
\end{table}

In the next case with $M=3$ and $a \gg 1$ (see Table \ref{owe3})
the structure of the triplet of the lowest and
almost degenerate
bound states $\psi^{(3)}_n(r)$ may be, similarly, characterized
by the absence of the nodal zero
in the ground-state wave function $\psi_0(r)$ [which is exact,
defined by Eq.~(\ref{431eq})],
and by the presence of one and two nodal zeros in the
first two excited states $\psi_1(r)$ and  $\psi_2(r)$,
respectively. This observation
is summarized in Table~\ref{owe3} where the
black dot $\bullet$ denotes a nodal zero of $\psi_n(r)$ and
its position on the real line of $r$. Schematically, the other two short-hand
symbols $\bigcap$ and
$\bigcup$ stand for the dominant gaussian component of
$\psi_n(r)$
with
positive and negative sign, respectively. In a self-explanatory manner
the other, smaller symbols represent the
exponentially suppressed parts of the curves.

\begin{table}[h]
\caption{Shapes and zeros of the $M-$plet of
$\psi_n^{(M)}(r)$
at $M=5$.
 }
\vspace{0.5cm}
 \label{owe5}
\centering {\small
\begin{tabular}{||c|c|c|c|c|c|c|c|c||}
\hline
     \hline
 $\bigcap$&&$\bigcap$&&$\bigcap$&
     &$\bigcap$&&$\bigcap$\\ \hline
 $\bigcap$&&$\bigcap$&&$\bullet$&
     &$\bigcup$&&$\bigcup$\\
     \hline
 $\bigcap$&&$\bigcap$&$\bullet$&$\bigcup$&$\bullet$
     &$\bigcap$&&$\bigcap$\\
     \hline
 $\bigcap$&$\bullet$&$\bigcup$&&$\bullet$&
     &$\bigcap$&$\bullet$&$\bigcup$\\
     \hline
 $\bigcap$&$\bullet$&$\bigcup$&$\bullet$&$\bigcap$&$\bullet$
     &$\bigcup$&$\bullet$&$\bigcap$\\
 \hline \hline
\end{tabular}}
\end{table}

%

What is certainly remarkable at $M=3$ is that
in contrast to the
preceding $M=2$ scenario
the spatial symmetry
leads,  in the first excited state,
to an almost complete suppression of the
central Gaussian component of the wave function.
We will see below that the same phenomenon also occurs in the analogous,
almost degenerate $M-$plets of the low-lying
bound states at any odd number of valleys $M$.
In this sense, the next, $M=4$
oscillation pattern as depicted
in Table~\ref{owe4} samples the simplest nontrivial
even-$M$
scenario which is less anomalous.

\begin{table}[h]
\caption{Shapes and zeros of the $M-$plet of
$\psi_n^{(M)}(r)$
at $M=6$.
 }
\vspace{0.5cm}
 \label{owe6}
\centering {\small
\begin{tabular}{||c|c|c|c|c|c|c|c|c|c|c||}
\hline
      \hline
  $\bigcap$&&$\bigcap$&&$\bigcap$&&$\bigcap$&&$\bigcap$&&$\bigcap$\\
  \hline
  $\bigcap$&&$\bigcap$&&$\bigcap$&$\bullet$&$\bigcup$&&$\bigcup$&&$\bigcup$\\
      \hline
  $\bigcap$&&$\bigcap$&$\bullet$&$\bigcup$&&$\bigcup$&$\bullet$&$\bigcap$&&$\bigcap$\\
  \hline
  $\bigcap$&$\bullet$&$\bigcup$&&$\bigcup$&$\bullet$&$\bigcap$
    &&$\bigcap$&$\bullet$&$\bigcup$\\
    \hline
  $\bigcap$&$\bullet$&$\bigcup$&$\bullet$&$\bigcap$&
     &$\bigcap$&$\bullet$&$\bigcup$&$\bullet$&$\bigcap$\\
  \hline
  $\bigcap$&$\bullet$&$\bigcup$&$\bullet$&$\bigcap$&$\bullet$
     &$\bigcup$&$\bullet$&$\bigcap$&$\bullet$&$\bigcup$\\
           \hline \hline
\end{tabular}}
\end{table}

A more or less routine extension of the pattern is displayed in Tables \ref{owe5}
and \ref{owe6} where we can see again the
slightly more involved but still regular wave-function-oscillation patterns
characterizing the distribution of the nodal zeros of
the $M-$plets of states $\psi_n(r)$ in dependence on the growth
of the excitation $n$.


\begin{table}[h]
\caption{Shapes and zeros of the $M-$plet of
$\psi_n^{(M)}(r)$
at $M=7$.
 }
\vspace{0.5cm}
 \label{owe7}
\centering {\small
\begin{tabular}{||c|c|c|c|c|c|c|c|c|c|c|c|c||}
\hline
     \hline
      $\bigcap$&&$\bigcap$&&$\bigcap$&&$\bigcap$&
     &$\bigcap$&&$\bigcap$&&$\bigcap$\\
     \hline
 $\bigcap$&&$\bigcap$&&$\bigcap$&&$\bullet$&
     &$\bigcup$&&$\bigcup$&&$\bigcup$\\
     \hline
 $\bigcap$&&$\bigcap$&&$\bigcap$&$\bullet$&$\bigcup$&$\bullet$
     &$\bigcap$&&$\bigcap$&&$\bigcap$\\
     \hline
 $\bigcap$&&$\bigcap$&$\bullet$&$\bigcup$&&$\bullet$&
     &$\bigcap$&$\bullet$&$\bigcup$&&$\bigcup$\\
     \hline
 $\bigcap$&$\bullet$&$\bigcup$&&$\bigcup$&$\bullet$&$\bigcap$&$\bullet$
     &$\bigcup$&&$\bigcup$&$\bullet$&$\bigcap$\\
      \hline
 $\bigcap$&$\bullet$&$\bigcup$&$\bullet$&$\bigcap$&&$\bullet$&
     &$\bigcup$&$\bullet$&$\bigcap$&$\bullet$&$\bigcup$\\
     \hline
 $\bigcap$&$\bullet$&$\bigcup$&$\bullet$&$\bigcap$&$\bullet$&$\bigcup$&$\bullet$
     &$\bigcap$&$\bullet$&$\bigcup$&$\bullet$&$\bigcap$\\
      \hline
      \hline
\end{tabular}}
\end{table}

In our last two Tables \ref{owe7}
and \ref{owe8} we are finally displaying
an extrapolation of the latter series of observations
to the next pair of systems with the six and seven
high and thick barriers, respectively.
Redundant as such an extrapolation might seem, we
are still displaying it because we believe that
it offers an insight in the general oscillation-theorem
pattern which is more intuitive and better understood than
its translation in the language of formulae.


\begin{table}[h]
\caption{Shapes and zeros of the $M-$plet of
$\psi_n^{(M)}(r)$
at $M=8$.
 }
\vspace{0.5cm}
 \label{owe8}
\centering {\small
\begin{tabular}{||c|c|c|c|c|c|c|c|c|c|c|c|c|c|c||}
\hline
      \hline
  $\bigcap$&&$\bigcap$&&$\bigcap$&&$\bigcap$&&$\bigcap$&&$\bigcap$&&$\bigcap$&&$\bigcap$\\
    \hline
  $\bigcap$&&$\bigcap$&&$\bigcap$&&$\bigcap$&$\bullet$&$\bigcup$&&$\bigcup$&&$\bigcup$&&$\bigcup$\\
        \hline
  $\bigcap$&&$\bigcap$&&$\bigcap$&$\bullet$&$\bigcup$
         &&$\bigcup$&$\bullet$&$\bigcap$&&$\bigcap$&&$\bigcap$\\
  \hline
  $\bigcap$&&$\bigcap$&$\bullet$&$\bigcup$&&$\bigcup$&$\bullet$&$\bigcap$
    &&$\bigcap$&$\bullet$&$\bigcup$&&$\bigcup$\\
        \hline
  $\bigcap$&$\bullet$&$\bigcup$&&$\bigcup$&$\bullet$&$\bigcap$&
     &$\bigcap$&$\bullet$&$\bigcup$&&$\bigcup$&$\bullet$&$\bigcap$\\
       \hline
  $\bigcap$&$\bullet$&$\bigcup$&$\bullet$&$\bigcap$&&$\bigcap$&$\bullet$
     &$\bigcup$&&$\bigcup$&$\bullet$&$\bigcap$&$\bullet$&$\bigcup$\\
    \hline
  $\bigcap$&$\bullet$&$\bigcup$&$\bullet$&$\bigcap$&$\bullet$&$\bigcup$&
     &$\bigcup$&$\bullet$&$\bigcap$&$\bullet$&$\bigcup$&$\bullet$&$\bigcap$\\
  \hline
  $\bigcap$&$\bullet$&$\bigcup$&$\bullet$&$\bigcap$&$\bullet$&$\bigcup$&$\bullet$
     &$\bigcap$&$\bullet$&$\bigcup$&$\bullet$&$\bigcap$&$\bullet$&$\bigcup$\\
                      \hline \hline
\end{tabular}}
\end{table}


\section{Discussion\label{discussion}}

\subsection{Multi-barrier models}

In the overall QES approach to Schr\"{o}dinger equations
a judicious reduction of the
variability of the potentials
is known to open the possibility of
obtaining certain {\em exact\,}
bound-state solutions
in a closed and compact {\em elementary-function\,} form.
For the specific class of polynomial potentials the discovery of
their QES property
dates back to the late seventies \cite{Singh}.
Treated, originally, as a mere mathematical curiosity,
the true impact of this approach
was enhanced by the subsequent developments
which revealed both the wealth of the
underlying mathematics \cite{Turbiner}
as well as the emergence of many
useful innovative implementations
of the QES idea in the various physical
contexts \cite{Ushveridze,Shifman}.

In our present paper we felt inspired
by the
success and
appeal of the QES idea going,
in its essence, against the conventional perception of
Schr\"{o}dinger equations.
In the long history of quantum mechanics the absolute
preference of the elementary
nature of $V(x)$ was always perceived as natural, dictated
also by the widespread belief in the heuristic
``principle of correspondence''
between the classical and quantum
laws of dynamics \cite{Messiah}.

In applications, the price to pay for such a purely
conventional preference was not too low.
Most of the constructions of
the states $\psi(x)$ and of their energies $E$
(i.e., after all, of the measurable effects)
had to be numerical
or, at best, perturbative \cite{Kato}.
On this background, the gain provided by the
QES-based trade-off appeared impressive.
Often, the
states $\psi(x)$ got simplified
at a very acceptable expense of
some formal constraints imposed upon $V(x)$.

In our present paper we
offered a new application of the QES philosophy
motivated by the need of description of the so called quantum catastrophes.
In a way complementing the results of our preceding paper \cite{arnoldium}
we paid main attention to the models in which the
potential is a composition of $M$ deep wells separated by an
$(M-1)-$plet of high barriers. At an arbitrary $M$,
our approach
may be compared with the more usual
implementation of the idea in which the QES
sextic-oscillator ground-state of formula (\ref{solu})
in section \ref{S2} is
generalized to acquire the form of the
product of such an exponential function
with a polynomial \cite{Singh}.

In the framework of our present project
of a search for the QES states with approximate degeneracy
we imagined that a new version of the more standard QES philosophy
could be based on
the construction of multiplets of the low-lying bound states
out of which just the lowest one is known exactly, while the
rest of the $M-$plet remains to be known just in
an approximate form. One can imagine a number of
directions of a possible further development of this
idea which have to be postponed to a future research.

\subsection{Excited states}

One of the characteristic features of our present multi-Gaussian
ground-state QES ansatzs is that
in the almost-impenetrable-barriers dynamical regime
the elementary
changes of the signs
attached to the separate Gaussians
could
form an alternative QES ansatz
yielding a
fairly precise description of the
first few excited-state wave functions $\psi_n(r)$ with $n=1,2,\ldots,M-1$.
In such an almost degenerate multiplet of states
the integer subscript $n$ characterizes
the degree of the excitation.
The systematics of the approximate sign-changing construction
was explained in the Tables
which clarified the relation between the parity of
the states and the admissible choices of the
signs of the separate Gaussians.

Such an overall result and remark must be complemented by an
observation that
the changes of signs of the separate Gaussian components
of $\psi_n(r)$
imply
the emergence of $n-$plets of the nodal zeros
in the wave function.
Their natural and generic localization
inside the range of the barriers may be interpreted
as a support of a tentative and
formal insertion
of $\psi_n(r)$
in the fundamental QES definition
(\ref{seb}) of the potential
(for the time being, let us denote the resulting
function with singularities by
the symbol $V_{(QES)}^{(n)}(r)$).
Although the emergence of these singularities
(i.e., poles) at the nodal zeros of  $\psi_n(r)$ would make
such potentials unacceptable,
they will still coincide, as functions of $r$, with
the original regular potential $V_{(QES)}^{(0)}(r)$ {\em locally},
i.e., more precisely, in the vicinities of the separate
harmonic-oscillator-mimicking
minima. Far from these minima
a regularization of the singularities
(i.e., of a mathematically correct description of
the tunneling) must be sought,
for example, via the analytic continuation techniques
\cite{sinde}.

In a preliminary test
of a hypothesis of a practical local coincidence of
singular $V_{(QES)}^{(n)}(r)$
with regular $V_{(QES)}^{(0)}(r)$
we performed a few numerical experiments. They
confirmed that not only the local but also the global
differences
$\Delta^{(n)}(r)=V_{(QES)}^{(n)}(r)-V_{(QES)}^{(0)}(r)$
cease to be small
just inside the classically forbidden subintervals of the coordinate.
Thus, these differences could be perceived, in some sense,
as a not too influential
or even, hopefully, systematically tractable perturbation.

At present,
unfortunately, the latter possibility
of an extension of the theory remains connected with only too
many open technical questions. Let us only add that
in some special cases
(characterized, e.g., by the use of a low-precision
computer arithmetics) the singularities of $\Delta^{(n)}(r)$
remained unnoticed even in some
numerical illustrative graphs of $V_{(QES)}^{(n)}(r)$.

\subsection{Catastrophe theory}

Occasionally, observable properties of
quantum systems are
described using
the
language of non-quantum physics
and catastrophe theory \cite{Thomchem}.
In our recent paper \cite{arnoldium}, in particular,
we choose the Arnold's
polynomial potentials
used as benchmarks in classical dynamical systems \cite{Arnold}.
Subsequently, we reinterpret them as models of dynamics of
one-dimensional quantum
systems.
In this setting the present formula (\ref{leading})
could be recalled as sampling
one of the key differences between
classical and quantum theory:
The quantum-equilibrium shift of energy
$\omega^{(\min)}>0$
would have to be zero in the
alternative classical-equilibrium context.
The
Thom's \cite{Thom}
purely geometric
classification
of the bifurcations of
classical equilibria would become inapplicable.
A consistent qualitative description of the
dynamics of
quantum equilibria
would necessarily have to be much subtler.

In our present paper we found the method of circumventing
one of the related serious though purely technical obstacles.
Our idea was twofold. Firstly, we emphasized that
in the quantum-theoretical analysis of equilibria it
makes sense to
restrict one's attention just to the ground-state solutions
of the underlying Schr\"{o}dinger equation and,
for one-dimensional systems,
to the mere ordinary differential Eq.~(\ref{se}).
Secondly, we imagined that such a physics-oriented restriction
can be perceived as mathematically represented by the
QES constructions.

The QES-non-QES differences still remained important, having
re-emerged on the global level.
The key advantage of the present approach
has been emphasized to
lie in a drastic simplification of the
predictions based on our exact knowledge of $\psi_{(QES)}(r)$.
In contrast to the conventional choice of polynomials $V_{(Arnold)}(r)$,
the price to pay for the availability of
the reconstructed (though still comparably elementary)
potentials $V_{(QES)}(r)$ lied only in their
slightly less user-friendly form of a
ratio of two polynomials.

From an
abstract mathematical
point of view such an upgrade could suffer
from a highly undesirable emergence of
the singularities in $V_{(QES)}(r)$.
The danger has been addressed here in some detail.
Two ways of its removal or suppression have been
emphasized. The first one was strictly physical:
Whenever one restricts attention just to
the relocalization bifurcations
in the ground-state regime, one cannot encounter any
singularities in $V_{(QES)}(r)$ because
the denominators (represented by
the
ground-state
ansatzs $\psi_{(QES)}(r)$) cannot support, by definition,
any nodal zeros.

From a second, slightly different point of view
we took into account the specific features
of the multi-well scenario
in the weak-tunneling dynamical regime.
In this limit
every potential barrier becomes high and thick
so that a few low-lying bound states start
forming a practically degenerate multiplet.
We indicated that in the nearest future
such a
strictly mathematical phenomenon might open,
after a suitable modification of the formalism,
the possibility
of an extension of the present ground-state QES constructions
to some of its regularized excited-state generalizations.

\section{Summary\label{summary}}

One of the roots of
the phenomenological appeal of one-dimensional multi-well
potentials $V(r)$
is in the contrast between their role in
classical and quantum physics.
In the former case, all of the eligible
equilibria are represented
by the
separate local minima of the potential.
This reduces any qualitative
prediction of dynamics
to a purely geometric study of
the  disappearance or confluence
of these minima \cite{revcat}.
In the quantum systems, in contrast,
the analysis of the situation becomes more complicated because
one always has to take into account the tunneling
which may spread the wave function
(and, hence, the measurable probability density)
over several, not necessarily equal local minima
of $V(r)$.

This being said
and properly taken into account, both the classical and quantum systems
share the possibility of a sudden change of their
state
after a comparably minor change of
the parameters in $V(r)$. This was advocated in \cite{arnoldium}
where, paradoxically,
several weak points of the conclusions
resulted from the purely technical decision of
working just with polynomial potentials.
This had two rather unpleasant consequences.
Firstly, it was not too easy to keep the shape of the
polynomial potential under control.
Secondly, even after we found and used an optimal
set of parameters and after we
managed to control the positions and values
of the minima of $V(r)$, it was not so easy to
construct also the corresponding measurable quantities
(i.e., energies or probability densities),
especially in the ``quantum catastrophe supporting''
dynamical regime
of an enhanced
sensitivity to the minor changes of the parameters.

In our present continuation of such an analysis
we described a new approach and picture of
quantum
catastrophes
based on a change of the underlying model-building paradigm.
The basic mathematical tool
of such a project has been found in
the quasi-exact philosophy of constructions of the systems
at a relocalization catastrophe instant.
We achieved a
thorough simplification of the
necessary construction
of the corresponding ground states $\psi_{(QES)}(r)$
via a QES-inspired
weakening of the traditional {\it a priori\,}
constraints imposed upon
the form of the effective interaction potentials.

The core of our message may be seen in
an explicit
description of
quantum systems admitting a
relocalization
catastrophe.
We managed to reach an optimal
balance between the combined
requirements of the mathematical feasibility and of a
phenomenological appeal of the result.
This means that in an unperturbed regime our
models are all set in
a highly unstable multi-centered ``cross-road'' state.
Its imminent evolution
is open to
several alternative processes of unfolding
under specific perturbations.

Having the resulting family of
models in which both $V(r)$
and $\psi(r)$
have an elementary
non-numerical form,
the details of
the evolution before and after the
passage of the system through the instant of
catastrophe remained out of the scope of the present
paper and are left to the reader.
Their study may be expected to proceed using the
standard methods of perturbation theory.
Still, in the language of mathematics
the exact solvability status of the system constructed
at a precise instant of the
relocalization
catastrophe
is rendered possible by the
QES approach in which
the state $\psi(r)$ is known exactly.

\subsection*{Acknowledgments}

The author acknowledges the financial support from the
Excellence project P\v{r}F UHK 2020.


\newpage

\begin{thebibliography}{00}

\bibitem{arnoldium}
M. Znojil, Ann. Phys. 413 (2020) 168050.

\bibitem{Zeeman}
E. C. Zeeman,
%
Catastrophe Theory-Selected Papers 1972-1977.
Addison-Wesley, Reading, 1977;

https://en.wikipedia.org/wiki/Catastrophe\_theory

\bibitem{Fluegge}
S. Fl\"{u}gge, Practical Quantum Mechanics I. Springer, Berlin,
1971.

\bibitem{Kato}
T. Kato, Perturbation theory for linear operators. Springer, Berlin, 1966.

\bibitem{EPappl}
M. Znojil,
J. Phys. A: Math. Theor.
41 (2008) 244027;

M. Znojil,
J. Phys. A: Math. Theor. 45 (2012) 444036;

P. Str\'{a}nsk\'{y}, M. Dvo\v{r}\'{a}k and P. Cejnar, Phys. Rev. E
97 (2018) 012112;



M. Znojil,
Phys. Rev. A 98 (2018) 032109.

\bibitem{2D3D}
M. Znojil, Ann. Phys. 416 (2020) 168161;

M. Znojil,
Mod. Phys. Lett. B 34 (2020)
2050378.

\bibitem{Ushveridze}
A. G. Ushveridze, Quasi-Exactly Solvable Models in Quantum Mechanics. IOPP, Bristol,
1994.
%
%

\bibitem{Turbiner}
A. V. Turbiner,
Comm. Math. Phys. 118 (1988)
467;

M. Znojil,
J. Phys. A: Math. Gen. 33 (2000) 4203;


M. Znojil,
J. Phys. A: Math. Gen. 33 (2000) 6825.





\bibitem{Shifman}
M. A. Shifman,
Int. J. Mod. Phys. A 4 (1989) 2897;

E. G. Kalnins, W. Miller Jr., G. S. Pogosyan,
J. Math. Phys. 47 (2003) 033502.

\bibitem{Turbinerb}
A. M. Ishkhanyan and G. L\'{e}vai,
 Phys. Scripta 95 (2020) 085202.

\bibitem{Turbinerc}
G. L\'{e}vai and J. M. Arias,
 Phys. Rev. C 81 (2010) 044304;
 
R. Budaca, P. Buganu, M. Chabab, A. Lahbas and M. Oulne,
 Ann. Phys. 375 (2016) 65.

\bibitem{Singh}
V. Singh, S. N. Biswas and K. Datta, Phys. Rev. D 18 (1978) 1901;

N. Saad, R. L. Hall and H. Ciftci,
J. Phys. A: Math. Theor. 39 (2006) 8477;

M. Znojil,
Phys. Lett. A 380 (2016) 1414;

G. L\'{e}vai and A. M. Ishkhanyan,
Mod. Phys. Lett. A 34 (2019) 1950134.

\bibitem{Arnold}
V. I. Arnold,
Catastrophe Theory. Springer-Verlag,  Berlin, 1992.

\bibitem{Thom}
R. Thom, Structural Stability and Morphogenesis: An Outline of a
General Theory of Models. Addison-Wesley, Reading,  1989.

\bibitem{Messiah}
A. Messiah, Quantum Mechanics I. North Holland, Amsterdam, 1961.

\bibitem{sinde}
P. Str\'{a}nsk\'{y}, M. \v{S}indelka, M. Kloc and P. Cejnar,
%
%
Phys. Rev. Lett. 125 (2020) 020401; 

P. Cejnar, P. Str\'{a}nsk\'{y}, M. Macek and M. Kloc,
Excited-state quantum phase transitions,
arXiv:2011.01662.
%

\bibitem{Thomchem}
X. Krokidis, S. Noury and B. Silvi, J. Phys. Chem 101 (1997) 7277;

W. Kirkby, Y. Yee, K. Shi and D. H. J. O'Dell,
Caustics in quantum many-body dynamics
(arxiv:2102.00288).


\bibitem{revcat}
J. Poston and I. Stewart, Catastrophe Theory and Its Applications.
Pitnam, London, 1978.

\end{thebibliography}
\end{document}